\documentclass[letterpaper, 10 pt, conference]{ieeeconf}  

\IEEEoverridecommandlockouts                              

\usepackage{graphics} 
\usepackage{epsfig} 
\usepackage{amsmath} 
\usepackage{amssymb,color}  
\usepackage{mdframed,url}

\title{\LARGE \bf
Partially Observable Games for Secure Autonomy*
}

\author{Mohamadreza Ahmadi$^{1}$, Arun A. Viswanathan$^{2}$, Michel D. Ingham$^{2}$, Kymie Tan$^{2}$, and Aaron D. Ames$^{1}$
\thanks{*The work described in this paper was performed at the California Institute of Technology, and at the Jet Propulsion Laboratory, California Institute of Technology, under a contract with the National Aeronautics and Space Administration (NASA).}
\thanks{$^{1}$M. Ahmadi and A. D. Ames are with the California Institute of Technology, 1200 E. California Blvd., Pasadena, CA 91125.
        {\tt\small [mrahmadi,ames]@caltech.edu}.}%
\thanks{$^{2}$A. Viswanathan, M. Ingham, and K. Tan are with the Jet Propulsion Laboratory, California Institute of Technology, 4800 Oak Grove Dr, Pasadena, CA 91109. 
        {\tt\small [arun.a.viswanathan, michel.d.ingham,kymie.tan]@jpl.nasa.gov}.}%
}
\newtheorem{definition}{\textbf{Definition}}

\DeclareMathOperator*{\argmax}{argmax}

\newcommand{\pr}{\ensuremath{\mathrm{Pr}}}

\newcommand{\N}{\mathbb{N}}

\newcommand{\Ireal}{[0,\, 1]\subseteq\mathbb{R}}  

\newcommand{\Distr}{\mathit{Distr}}

\newcommand{\distDom}{X}

\newcommand{\distFunc}{\mu}
\newcommand{\distDomElem}{x}
\DeclareMathOperator{\supp}{supp}

\DeclareMathOperator{\dom}{dom}

\newcommand{\sinit}{s_{\mathrm{I}}} 

\newcommand{\mdp}{M}

\newcommand{\probmdp}{\mathcal{P}}

\newcommand{\actionMap}{\ensuremath{\mathcal{\gamma}}}
\newcommand{\nodeTransition}{\ensuremath{\mathcal{\delta}}}

\newcommand{\ObsSym}{{Z}}
\newcommand{\ObsFun}{{O}}
\newcommand{\obs}{\ensuremath{z}}

\newcommand{\posg}{\mathcal{G}}

\newcommand{\spOne}{\ensuremath{S_{\circ}}}
\newcommand{\spTwo}{\ensuremath{S_{\Box}}}

\newcommand{\sg}{\ensuremath{G}}  

\newcommand{\states}{\ensuremath{S}}

\newcommand{\sched}{\ensuremath{\sigma}}

\newcommand{\osched}{\ensuremath{\mathit{\sigma}}}

\newcommand{\Act}{\ensuremath{\mathit{Act}}}
\newcommand{\act}{\ensuremath{a}}

\newcommand{\pathset}{\mathsf{Paths}}

\newcommand{\pathsfin}{\pathset_{\mathit{fin}}}

\newcommand{\last}[1]{\mathrm{last}(#1)}

\DeclareMathAlphabet{\mathpzc}{OT1}{pzc}{m}{it}
\def\presuper#1#2%
  {\mathop{}%
   \mathopen{\vphantom{#2}}^{#1}%
   \kern-\scriptspace%
   #2}

\usepackage[most]{tcolorbox}

\begin{document}

\maketitle
\thispagestyle{empty}
\pagestyle{empty}

\begin{abstract}
Technology development efforts in autonomy and cyber-defense have been evolving independently of each other, over the past decade.
 In this paper, we report our ongoing effort to integrate these two presently distinct areas into a single framework. To this end, we propose the two-player partially observable stochastic game formalism to capture both high-level autonomous mission planning under uncertainty and adversarial decision making subject to imperfect information. We show that synthesizing sub-optimal strategies for such games is possible under finite-memory assumptions for both the autonomous decision maker and the cyber-adversary. We then describe an experimental testbed to evaluate the efficacy of the proposed framework.
\end{abstract}

\section{INTRODUCTION}

The growing ubiquity of  autonomous systems, their use in ever more remote and unknown environments, and the increasing sophistication of cyber threats are driving a need for unprecedented system resilience, coupling robust autonomy with efficient cyber-defense strategies~\cite{kott2018intelligent,falco2018vacuum}.
Consider the push to develop swarms of smallsats in low Earth orbit. 
Cost-effective operations of such
swarms require improved autonomy capabilities, both onboard and on the ground. However, complex
autonomous behavior makes such systems susceptible to malicious tampering. 
Similarly, current unmanned
air/ground/underwater systems
rely on various signals
for communication and localization and are already vulnerable to spoofing attacks.
A GPS spoofing attack against such systems could result in malicious GPS coordinates being fed to the vehicle, causing it to be (mis)guided on an adversary's behest~\cite{giray2013anatomy}.
A resilient autonomous system should be able to detect attacks against itself, diagnose the probable causes, and automatically take corrective actions while ensuring the system's low/high-level goals and objectives are achieved. 

However, a primary challenge to achieving this vision of integrated cyber and physical resilience is that technology development efforts in autonomy and cyber-defense are presently evolving independently of each other. Our work aims to reverse this trend. 
Our overall goal is to develop and demonstrate resilient autonomy for  autonomous agents, by extending existing risk-aware planning and execution capabilities~\cite{mcghan2016} with a combination of state-of-the-art model-based reasoning for situational and self-awareness and active cyber-defense mechanisms. 

Current cyber adversaries can study the defender's behavior, identify security caveats, and modify their actions adaptively~\cite{tankard2011advanced}. 
To tackle these security challenges, cyber-agents require adversarial decision making under uncertainty. Furthermore, agents cannot directly observe their adversary's true state and/or intention. Hence, active cyber-defense methods necessitate dealing with partial observations~\cite{ahmadi2019risk} and imperfect/incomplete information. 
A game-theoretic framework known as partially observable stochastic games~(POSG)~\cite{kumar2009dynamic} provides a promising mathematical formalism for these capabilities. 

In this paper, we report our preliminary methodology based on POSGs to integrate high-level autonomy and adversarial decision making. Our method based on POSGs is aimed at addressing cyber-physical threats caused by active cyber-adversaries, for example, as seen in the Stuxnet attack~\cite{langner2011stuxnet}, wherein the attacker modifies their strategy in reaction to defensive actions. We show that the solution to the POSG can be cast as an optimization problem. Then, we propose an experimental setup to evaluate our technique. In summary, we hope to make the following contributions:
\begin{itemize}
    \item Novel high-level resilient autonomy in the presence of active cyber-attacks leveraging the POSG framework;
    \item Demonstration of an integrated "defense-in-depth" capability for secure autonomy of cyber-physical systems.
\end{itemize}

The rest of the paper is organized as follows:
Section~\ref{sec:background} discusses the threat model for a cyber-physical system such as an UAV, an autonomous robot, or a swarms of spacecrafts;
Section~\ref{sec:methodology} discusses our proposed methodology using POSGs; Section~IV discusses our experimental evaluation methodology followed by our conclusions and future work in Section~V.

\begin{figure}[t!]\label{fig:cast}
\begin{center}
    \includegraphics[width=7.5cm]{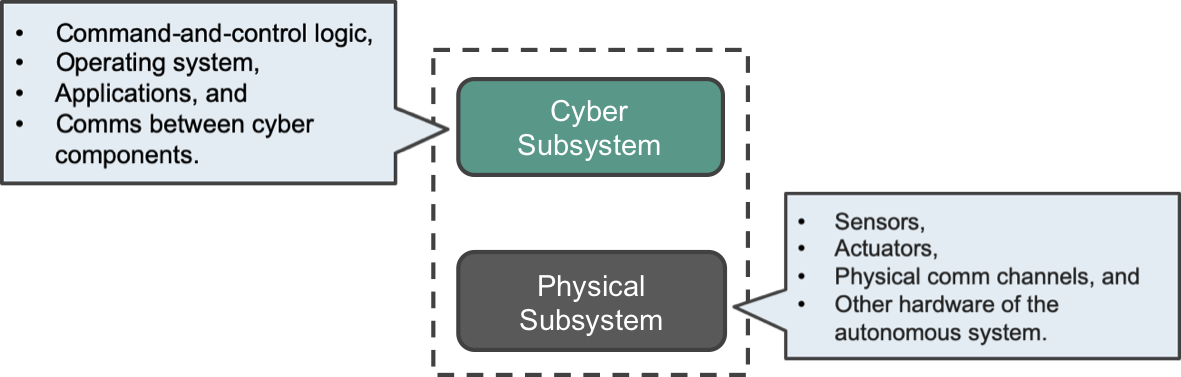}
\end{center}
\caption{A simplified model of an autonomous system.}
\label{fig:cpmodel}
\end{figure}

\section{Cyber-Physical Threat Model}
\label{sec:background}

In this section, we first describe a model of an autonomous system, followed by a description of adversarial goals and a high-level taxonomy of threats.

Figure~\ref{fig:cpmodel} shows a simplified model of an autonomous system (agent), 
containing two subsystems: \emph{cyber} and \emph{physical}.
The cyber subsystem encapsulates functionality such as command and control logic, operating system, applications and any communications between the cyber components.
Cyber components may be located on the agent or be external to the agent.
Multi-agent systems may have a centralized cyber subsystem coordinating the agents. 
The physical subsystem encapsulates entities such as sensors, actuators, physical communication channels, and any other hardware comprising the autonomous system.
\begin{figure}[t!]\label{fig:cast}
\begin{center}
    \includegraphics[width=7.5cm]{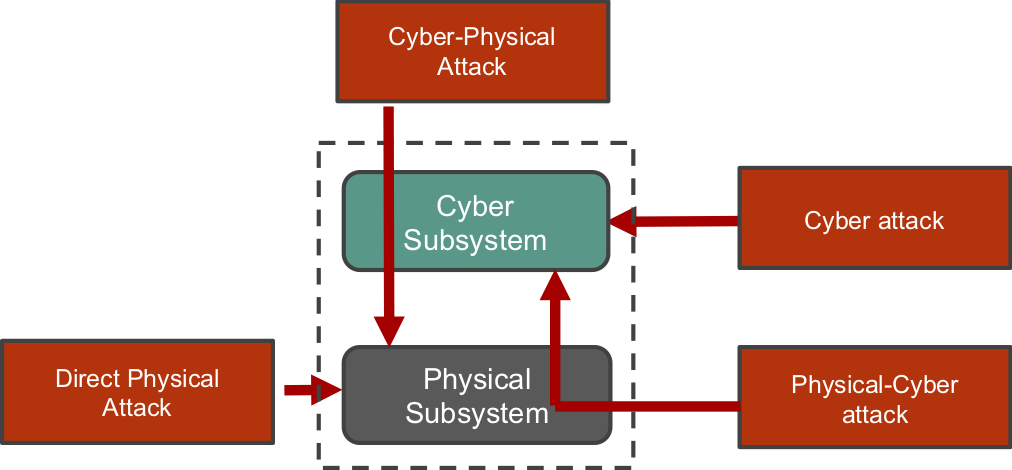}
\end{center}
\caption{Cyber-Physical Threat Model}
\label{fig:threatmodel}
\end{figure}
An attacker would want to gain malicious control, cause damage, or deny service to prevent the autonomous system from achieving its goals.
Referring to Figure~\ref{fig:threatmodel}, there are four different kinds of attacks an adversary could use to achieve their goals.
\begin{description}
\item[\textbf{Cyber Attack}] \hfill \\ 
    A cyber attack directly targets the components in the cyber subsystem. For example, a denial-of-service attack against the communication network of an autonomous system is an example of a cyber attack. 
\item[\textbf{Physical Attack}] \hfill \\ 
    A physical attack targets the components in the physical subsystem. For example, a ballistic impact is a type of physical attack which could damage physical components of an autonomous system. A physical attack often requires physical proximity to the system.
\item[\textbf{Cyber-Physical Attack}] \hfill \\ 
    In a cyber-physical attack, an attacker leverages a cyber vulnerability with the intent to affect the physical subsystem. For example, malicious input injection attacks such as the malicious command or data injection seen in recent car hacks~\cite{wiredcarhack2015}. 
    Cyber-physical attacks are often the most devastating as they can be initiated remotely, and cause serious damage to the physical subsystem.
\item[\textbf{Physical-Cyber Attack}] \hfill \\ 
    In a physical-cyber attack, an attacker influences the cyber subsystem by attacking the components in the physical subsystem. For example, an attack on the physical sensors of an autonomous system (say the IMU), may cause inaccurate data to be sent upstream to the cyber components (for example, incorrect location information), thereby causing incorrect decision-making and response by the cyber component. 
\end{description}

In our work, we focus on the cyber-physical and physical-cyber kinds of attacks, as these 
attacks cross boundaries and as such, are often more subtle and difficult to diagnose, and consequently pose significant risk to missions.
In addition, existing cyber or physical defenses generally do not protect against these attacks. 

In the next section, we describe a mathematical formalism considering cyber-physical and physical-cyber attacks. 

\section{Methodology:  Two-Player POSG}
\label{sec:methodology}

A POSG is formally defined as follows.

A \emph{probability distribution} over a finite or countably infinite set $\distDom$
is a function $\distFunc\colon\distDom\rightarrow\Ireal$ with $\sum_{\distDomElem\in\distDom}\distFunc(\distDomElem)=\distFunc(\distDom)=1$.
The set of all distributions on $\distDom$ is $\Distr(\distDom)$. The support of a distribution $\distFunc$ is
$\supp(\distFunc) = \{x\in\distDom\,|\,\distFunc(x)>0\}$.
A distribution is \emph{Dirac} if $|\!\supp(\distFunc)| = 1$. 


\begin{definition}
  \label{def:sg}
 \textit{ A \emph{stochastic game} (SG) is a tuple $G=(S,s_I,Act,\mathcal{P})$ with a finite set $\states=\spOne\cup\spTwo$ of \emph{states}, a set $\spOne$ of Player~1 states, a set $\spTwo$ of Player~2 states, the \emph{initial state} $\sinit\in\states$, a finite set $\Act=\Act_\circ \cup \Act_\square$ of \emph{actions}, and a \emph{transition function} 
  $\probmdp\colon \states\times\Act\rightarrow\Distr(\states)$. We define costs using a state-action cost function $C : S \times \Act \rightarrow \mathbb{R}_{\geq 0}$.}
\end{definition}

%

A Markov decision process (MDP) is an SG in which $\spOne = \emptyset$, and consequently $S = \spTwo$. A \emph{path} of an SG $\sg$ is an (in)finite sequence $\pi = s_0\xrightarrow{\act_0}s_1\xrightarrow{\act_1} ~s$,
where $s_0=\sinit$, $s_i\in\states$, $\act_i\in\Act$, and $\probmdp(s_i,\act_i)\neq 0$ for all $i\in\N$.
For finite $\pi$, $\last{\pi}$ denotes the last state of $\pi$. 
The set of (in)finite paths of $\sg$ is $\pathsfin^{\sg}$ ($\pathset^{\sg}$).

To define a probability measure over the paths of an SG $G$, the non-determinism needs to be
resolved by \emph{strategies}.

\begin{definition}[SG strategy]
  \label{def:strategy}
\textit{   A \emph{strategy} $\sched$ for $\sg$ is a pair $\sched = (\sched_\circ,\sched_\square)$ of functions
  $\sched_i\colon \{\pi\in \pathsfin^{\sg} \mid\last{\pi}\in S_i\}\to\Distr(\Act)$ such that for all $\pi\in \pathsfin^{\sg}$,
  $\{ \act\mid\sched_i(\pi)(\act)>0\} \subseteq \Act$, $i\in\{\circ,\square\}$. }
\end{definition}
A Player-$i$ strategy $\sched_i$ (for $i\in\{\circ,\square\}$) is \emph{memoryless} if $\last{\pi}=\last{\pi'}$ implies $\sched_i(\pi)=\sched_i(\pi')$ for all $\pi,\pi'\in\dom(\sched_i)$. 
It is \emph{deterministic} if $\sched_i(\pi)$ is a Dirac distribution for all $\pi\in\dom(\sched_i)$. 

A strategy~$\sigma$ for an SG resolves all non-deterministic choices, yielding an \emph{induced MC}, for which a \emph{probability measure} over the set of infinite paths is defined by the standard cylinder set construction~\cite{BK08}.
These notions are analogous for MDPs.

In our framework, $\spOne$ consists of the physical and mission states,  e.g.  robot(s)  location  and obstacles, or the autonomous decision maker; whereas, $\spTwo$ corresponds to the internal states of the cyber-adversary. These states are not directly observable to either player; the players must infer the probability of their opponent being at different states based on the observations received at every step of the game. Thus, we have a POSG as follows (see Figure~3).

\begin{figure}[t]\label{fig:posg}
\begin{center}
    \vspace{-.5cm}
    \includegraphics[width=6cm]{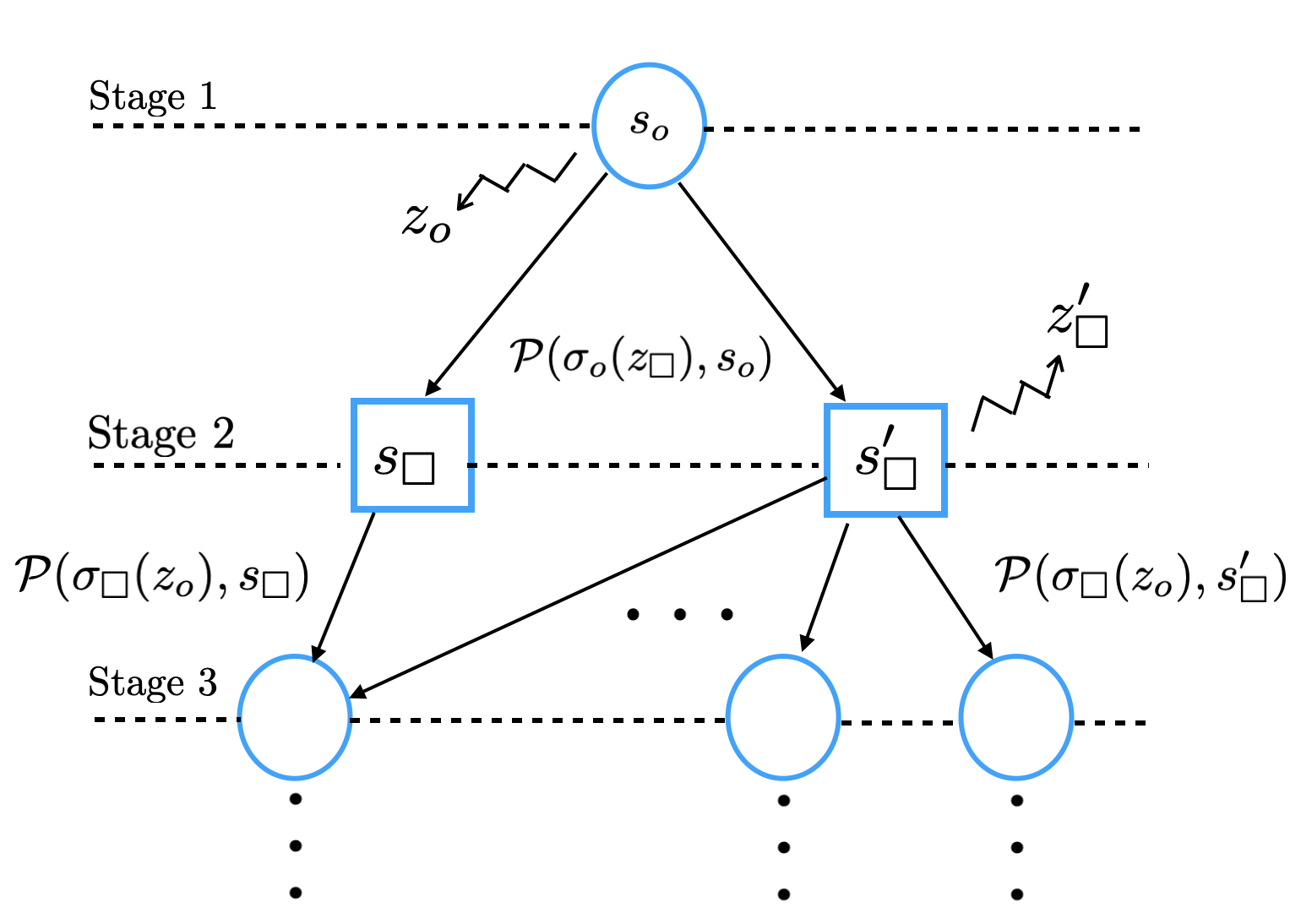}
    \end{center}
    \vspace{-.5cm}
    \caption{Three stages of an example POSG.  The states of the players need to be estimated based on the observations, and in the case of the attacker $\square$, counteracted. The game starts at $s_o$ with an initial observation $z_\square$.}
    \end{figure}

\begin{definition}
  \label{def:pomdp}
\textit{   A \emph{partially observable stochastic game (POSG)} is a tuple $\mathcal{G}=(G,Z_\circ,Z_\square,O_\circ,O_\square)$, with $G=(S,s_I,Act,\mathcal{P})$ the \emph{underlying SG of $\posg$}, $\ObsSym_\circ$ and $\ObsSym_\square$ are finite set of observations for Player 1 and 2, respectively, and $\ObsFun_\circ \colon S\rightarrow\ObsSym_\circ$ ($\ObsFun_\square \colon S\rightarrow\ObsSym_\square$) the \emph{observation function} for Player~1 (Player 2).}
\end{definition}



We lift the observation function to paths: For $\pi=s_0\xrightarrow{\act_0} s_1\xrightarrow{\act_1} ~ s_n\in\pathsfin^{\mdp}$, the associated \emph{observation sequence} is $\ObsFun(\pi)=\ObsFun(s_0)\xrightarrow{\act_0} \ObsFun(s_1)\xrightarrow{\act_1} ~\ObsFun(s_n)$.

\begin{definition}[POSG Strategy]
  \label{def:obsstrategy}
\textit{   An \emph{observation-based strategy} $\osched_i$ for Player $i$ in  POSG $\posg$ is a strategy $\osched_i$ for Player $i$ in the underlying SG $\sg$ such that $\osched_i(\pi)=\osched_i(\pi')$ for all $\pi,\pi'\in\pathsfin^{\posg}$
  with $\ObsFun_i(\pi)=\ObsFun_i(\pi')$. }
\end{definition}

Applying the strategy $\osched=(\osched_\circ,\osched_\square)$ to a POSG $\posg$ resolves all nondeterminism and partial observability, resulting in the \emph{induced Markov chain} $\posg^\osched$.

However, since POSGs simply extend POMDPs to multiple players, computing optimal strategies requires infinite memory~\cite{chatterjee2016decidable}. To circumvent this difficulty, we represent observation-based strategies with \emph{finite} memory and we use \emph{finite-state strategies} (FSSs) (see also FSSs in Delay Games~\cite{winter2019finite}).  If such an FSS has $n$ memory states, we say the  memory size for the underlying strategy $\osched$ is $n$.

\begin{definition}[FSS]
  \label{def:fsc}
\textit{   A \emph{finite-state strategy (FSS)} for Player $i$ in POSG $\posg$ is a tuple $\mathcal{A}_i=(N_i, n^I_i,\gamma_i,\delta_i)$, where $N_i$ is a finite set of \emph{memory states},
  $n^I_i\in N_i$ is the \emph{initial memory state}, $\gamma_i$ is the \emph{action mapping} $\gamma_i\colon N_i\times\ObsSym_i\rightarrow\Distr(\Act)$,
  and $\nodeTransition_i$ is the \emph{memory update} $\nodeTransition_i\colon N_i\times\ObsSym_i\times\Act\rightarrow \Distr(N_i)$.
  The set $FSS_k^{\posg}$ denotes the set of FSSs with $k$ memory states, called \emph{$k$-FSS}s.}
\end{definition}

At each stage of the game, for each player, from a node $n$ and the observation $\obs$ in the current state of the POSG, the next action $\act$ is chosen from $\Act(\obs)$ randomly as given by $\actionMap(n, \obs)$. Then, the successor node of the FSS is determined randomly via $\nodeTransition(n, \obs, \act)$.


\subsubsection*{\textbf{A POSG for Secure Autonomy}}

With the FSS assumption, the goal is then to maximize the probability of satisfying mission specifications, e.g. reach goal region while avoiding obstacles in the presence of cyber-adversarial activity. Next, we formally define the game objective.

\emph{Game Objective:}
For a POSG $\posg$ and a mission specification defined by a temporal logic formula $\varphi$, we consider the  \emph{probability} $\pr^\posg(\varphi)$  to satisfy $\varphi$.

The specification $\varphi$ is satisfied for a strategy $\osched=(\osched_\circ,\osched_\square)$ and the POSG $\posg$ with probability $\lambda \in [0,1]$, if the probability $\pr^{\posg^\osched}(\varphi) = \lambda$ or simply if the induced Markov chain by applying strategy $\sigma$ satisfies the specification with probability~$\lambda$. At this point, we have the following game formulation of secure autonomy problem.

\begin{mdframed}[backgroundcolor=gray!50!white]\label{prob1}
\textbf{Problem~1}: \textit{Given a POSG  $\mathcal{G}=(G,Z_\circ,Z_\square,O_\circ,O_\square)$, mission specification defined by a temporal logic formula $\varphi$, memory bounds $n_\circ$ for the decision maker and $n_\square$ for the cyber-adversary, compute a FSS $\sigma^*_\circ$ such that}
\begin{equation*}
    \sigma^*_\circ = \argmax_{\sigma_\circ \in FSS_{n_\circ}^{\posg}} \min_{\sigma_\square \in FSS_{n_\square}^{\posg}} ~\pr^{\posg^\osched}(\varphi).
\end{equation*}
\end{mdframed}

 In Problem 1, we look for worst-case resilient strategies such that the probability of satisfying the specifications is maximized. Alternatively, we can search for resilient strategies that maximize the expected value of meeting the specifications in the presence of adversarial activity. Indeed, we can approximate $\pr^{\posg^\osched}(\varphi)$ with an expected total cost type constraint~\cite{ASB20}. Then, for reachability type formulae such as $\varphi = \diamond T$ (eventually reach a goal region represented by the states in $T$), where $T \subset S$. The solution to Problem~1 can be found by solving an optimization problem as follows (see~\cite{ACJJKT19} for the derivation for one-sided POSGs).

For $s_{\circ}\in \spOne,$ and $s_{\Box} \in \spTwo $, we define the \emph{cost variables} $c_{s_{\circ}}\geq 0$ $c_{s_{\Box}}\geq 0$ that represent the expected cost of reaching $T\subseteq S$ with 
$c_{s_I}$ being the expected cost of reaching to $T$ from the initial state $s_I$.  Let $\gamma \in [0,1)$ be the discount factor to ensure finite total expected cost. We then have the optimization problem:
				\begin{align}
		& \underset{c_{s_{\circ}},\sigma_\circ }{\text{minimize}}~\underset{c_{s_{\Box}},\sigma_\square}{\text{maximize}} \quad c_{\sinit}\label{eq:min_sg_det}\\
			&\text{subject to} \nonumber \\
		&	 c_s=0, \quad \forall s\in T,&\label{eq:targetprob_sg_det}\\
				&	 \sum_{\act\in \Act_\circ}\sched^\obs_\act=1,\quad	\forall \obs\in \ObsSym_\square,&\label{eq:well-defined_probs_sg_det}\\
								&	 \sum_{\act\in \Act_\square}\sched^\obs_\act=1,\quad	\forall \obs\in \ObsSym_\circ,&\label{eq:well-defined_probs_sg_det2}\\
								& c_{s_{\Box}} =  C(s_{\Box},\act)+\gamma \sum_{\act\in\Act_\circ}\sched^{\ObsFun(s_{\Box})}_\act  ~ \sum_{s'_{\circ}\in S_{\circ}}	\probmdp(s_{\Box},\act,s'_{\circ}) ~ c_{s'_{\circ}},\nonumber\\
								&\qquad\qquad\qquad\qquad	\forall s_{\Box}\in S_{\Box}\setminus T,\,\forall \sched^{\ObsFun(s_{\Box})}_\act \in \sched_\square, \label{eq:probcomputation_sg1_det}	\\ 
 & c_{s_{\circ}} = C(s_{\circ},\act)+ \gamma  \sum_{\act\in\Act_\square}\sched^{\ObsFun(s_{\circ})}_\act \sum_{s'_{\Box}\in S_{\Box}}\probmdp(s_{\circ},\act,s'_{\Box}) ~ c_{s'_{\Box}}, \nonumber \\&\qquad\qquad\forall s_{\circ}\in S_{\circ}\setminus T, \,\forall \sched^{\ObsFun(s_{\circ})}_\act \in \sched_\circ.	\label{eq:probcomputation_sg3_det}
		\end{align}	
		
The objective in~\eqref{eq:min_sg_det} implies the decision maker $\circ$ is minimizing the cost of reaching $T$ from the initial state; whereas, the cyber-adversary $\square$ is trying the maximize the cost. We assign the expected cost of the states in the target set $T$ to 0 by the constraints in~\eqref{eq:targetprob_sg_det}. We ensure that the strategies of the decision maker and the cyber-adversary are well-defined with the constraints in~\eqref{eq:well-defined_probs_sg_det} and~\eqref{eq:well-defined_probs_sg_det2}. The constraints in~\eqref{eq:probcomputation_sg1_det}--\eqref{eq:probcomputation_sg3_det} give the computation for the expected cost in the states of the POSG via dynamic programming.

We will develop methods based on heuristics and nonlinear programming to solve the resultant POSGs algorithmically and we will study trade-offs between resilience (cyber side) and mission goals (physical side). 
Preliminary work in solving POSGs was carried out in~\cite{ACJJKT19} for the case when only the  adversary is subject to partial observation with application to network security. Instead of solving the full game, we used model checking to synthesize a set of strong (sub-optimal) strategies for the adversary and then composed robust defensive strategies. 

\section{Experimental Evaluation}

\begin{figure}[t]\label{fig:cast}
\begin{center}
    \includegraphics[width=7cm]{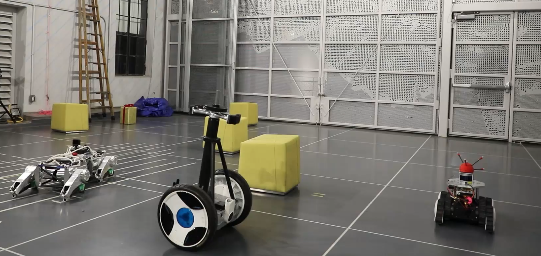}
    \end{center}
    \caption{Three robots involved in experimental evaluations at CAST: (left) quadruped, (center) Segway, and (right)  Flipper.}
    \end{figure}

The efficacy of the developed methods will be evaluated through experiments with three autonomous agents (a Segway, a quadruped, and a Flipper robot) in Caltech's Center for Autonomous Systems and Technologies (CAST) as depicted in Figure~4. The quadruped and the Flipper robot will be tasked to locate the target and the obstacles, respectively; whereas, the Segway is able to retrieve the target once the quadruped and Flipper explore the area. Flipper is equipped with a 3D LIDAR and a router. The quadruped robot is equipped with a high-resolution camera, an Inertial Measurement Unit (IMU), and a router. The Segway only has wheel odometry, an IMU, and a router. The centralized decision making is carried out through a computer connected to the robots via a wifi network. The sensor signals of each robot are also sent back to the computer via the same network. 

Our previous experiments in this setting were concerned with safe autonomy enforced by discrete-time barrier functions~\cite{2019arXiv190307823A}, i.e., in the absence of cyber-adversaries (watch the experimental demonstration at~\cite{youtube}).

The goal of our next set of experiments is to find and retrieve the target in the presence of cyber adversarial activity. This experimental setup is described next.

The states of the POSG for Player $\circ$ (the decision-maker) correspond to the locations of each agent, obstacles, and the goal. The actions for Player $\circ$ include moving $Left, Right, Up, Down$ for each agent.  The two states of Player $\square$ (cyber-adversary) are $Quadruped, Flipper$ corresponding to the two surveying agents. The actions of the attacker are to $Take Down$ or $Wait$. If $Take Down$ is chosen at one stage of the game, for example, for the Flipper robot, the robot will not move in the next step and its observation cannot be used for path planning. On the other hand, $Wait$ means no action is taken by the adversary. 

The objective of Player $\circ$ is then to maximize the probability of retrieving the target and avoiding obstacles; whereas, the Player $\square$ attempts to minimize this probability. This POSG fits in the framework of Section~II and can be used to assure high-level mission autonomy as well as cyber-resilience. 
This initial abstract problem formulation will provide a basis for more realistic (high-fidelity) solutions to the real-world problem in future work, e.g., examining real injected cyber-attacks and practical defensive responses.

\section{CONCLUSIONS}

We described our ongoing research on the fusion of autonomous decision making and active cyber-resilience. We proposed a POSG that can capture high-level mission specifications, uncertainty, partial observation, and adversarial decision making. Although finding optimal strategies for POSGs is undecidable, we discussed finite-memory strategies as computationally tractable alternatives. Finally, we presented an experimental testbed, methodology and a case study to evaluate our secure autonomy techniques in the future.  

\addtolength{\textheight}{-12cm}   





\section*{ACKNOWLEDGMENT}



The authors thank Prof. Richard M. Murray at Caltech and Dr. Nils Jansen at the Radboud University Nijmegen.

\bibliographystyle{plain}
{\bibliography{refs}}

\end{document}